\documentclass[twocolumn,english,notitlepage,showpacs,superscriptaddress]{revtex4-1}
\usepackage[T1]{fontenc}
\usepackage[latin9]{inputenc}
\usepackage{float}
\usepackage{amstext}
\usepackage{graphicx}
\usepackage{esint}
\usepackage[colorlinks,citecolor=blue,linkcolor=red]{hyperref}
\usepackage{color,soul}
\usepackage{comment}
\makeatletter
\usepackage{babel}

\makeatother

\usepackage{babel}
\begin{document}

\title{Universal constraint for efficiency and power of a low-dissipation heat engine}

\author{Yu-Han Ma}

\affiliation{Beijing Computational Science Research Center, Beijing 100193, China}

\affiliation{Graduate School of Chinese Academy of Engineering Physics, Beijing
100084, China}

\author{Dazhi Xu}
\email{dzxu@bit.edu.cn}

\selectlanguage{english}%

\affiliation{Center for Quantum Technology Research and School of Physics, Beijing
Institute of Technology, Beijing 100081, China}
\affiliation{Graduate School of Chinese Academy of Engineering Physics, Beijing
100084, China}

\author{Hui Dong}
\email{hdong@gscaep.ac.cn}

\selectlanguage{english}%

\affiliation{Graduate School of Chinese Academy of Engineering Physics, Beijing
100084, China}

\author{Chang-Pu Sun}
\email{cpsun@csrc.ac.cn}

\selectlanguage{english}%

\affiliation{Beijing Computational Science Research Center, Beijing 100193, China}

\affiliation{Graduate School of Chinese Academy of Engineering Physics, Beijing
100084, China}
\begin{abstract}
The constraint relation for efficiency and power is crucial to design
optimal heat engines operating within finite time. We find a universal
constraint between efficiency and output power for heat engines operating in the low-dissipation regime. Such constraint
is validated with an example of Carnot-like engine. Its microscopic
dynamics is governed by the master equation. Based on the master equation,
we connect the microscopic coupling strengths to the generic parameters
in the phenomenological model. We find the usual assumption of low-dissipation is achieved when the coupling to thermal environments is stronger than the driving speed.
Additionally, such connection allows the design of practical cycle to
optimize the engine performance. 
\end{abstract}

\pacs{to be added later}
\maketitle

\section{Introduction}

For a heat engine, efficiency and power are the two key quantities
to evaluate its performance during converting heat into useful work.
To achieve high efficiency, one has to operate the engine in a nearly
reversible way to avoid irreversible entropy generation. In thermodynamic textbook, Carnot cycle
is an extreme example of such manner, with which the fundamental upper
bound of efficiency $\eta_{\mathrm{C}}=1-T_{c}/T_{h}$ is achieved
with infinite long operation time \cite{carnot}. Such long time reduces
the output power, which is defined as converted work over operation
time. Generally, efficiency reduces as power increase, or vice visa.
Such constraint relationship between efficiency and power is critical
to design optimal heat engines. Attempts on finding such constraint
are initialized by Curzon and Ahlborn with a general derivation of
the efficiency at the maximum power (EMP) $\eta_{\mathrm{CA}}^{\mathrm{EMP}}=1-\sqrt{T_{c}/T_{h}}$
\cite{CA,CA1,CA2}. The EMP of heat engine has attracted much attention
and has been studied by different approaches in theory, such as Onsager relation \cite{EMP3,EMP4,Udo}
and stochastic thermodynamics \cite{EMP5,EMP6} with various systems
\cite{EMP7,EMP8,EMP9,EMP11,EMP12}; and in experiment
\cite{Brownian2016, Singer2016}. Esposito et.~al. discussed the
low-dissipation Carnot heat engine by introducing the assumption that
the irreversible entropy production of finite-time isothermal
process is inversely proportional to time \cite{key-1-esposito},
and obtained a universal result of the upper and lower bounds of the
EMP via optimization of the dissipation parameters.

\begin{figure}[t]
\includegraphics[width=8.5cm]{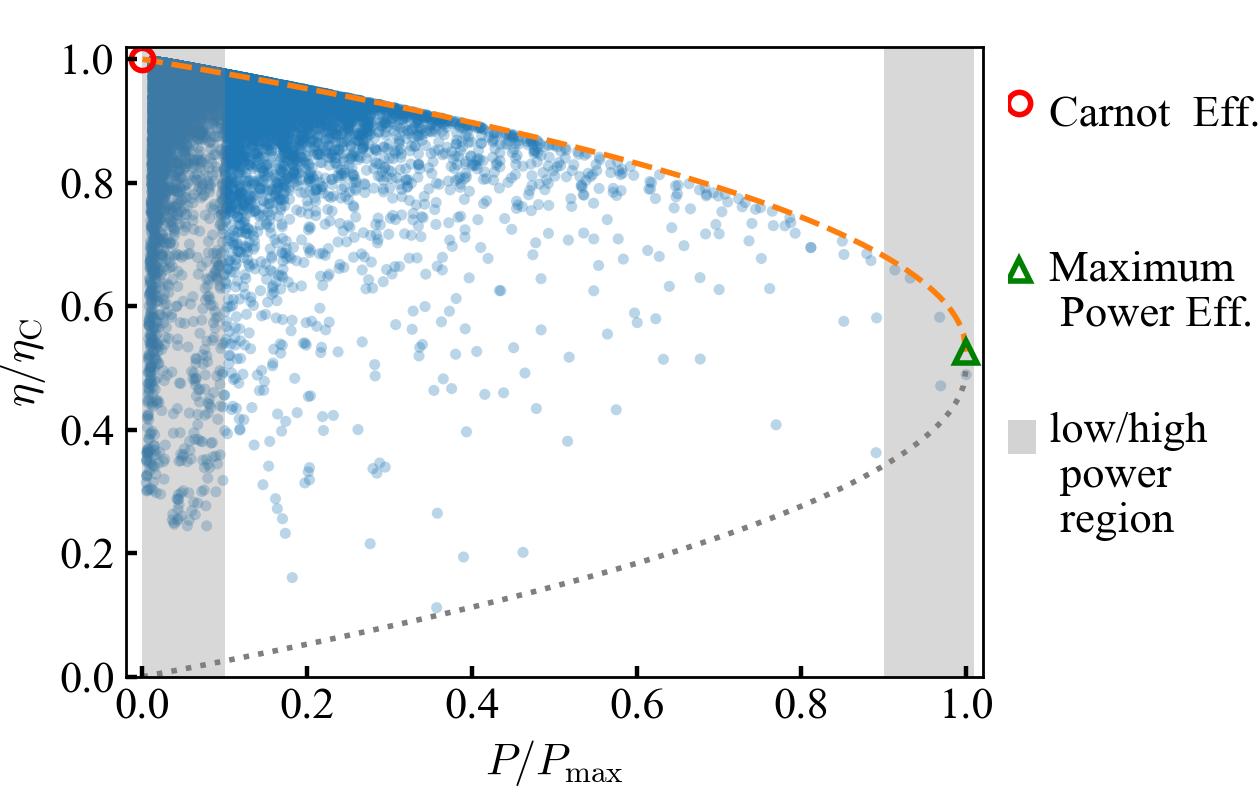}

\caption{(Color online) Constraint on normalized efficiency $\widetilde{\eta}\equiv\eta/\eta_{\mathrm{C}}$
and output power $\widetilde{P}\equiv P/P_{\mathrm{max}}$. The orange
curve shows the constraint relation of Eq.~(\ref{eq:constraint}).
Dots show the normalized efficiency and output power of a simple two-level
atomic heat engine. The gray dotted curve shows the lower bound,
which will be derived as in later discussion. The red circle denotes
the Carnot efficiency $\eta_{\mathrm{C}}$, the green triangle marks
the maximum power efficiency obtained in Ref.~\cite{key-1-esposito}.
The gray area represents the bound derived in Ref.~\cite{constriant2-1}.}

\label{figure1} 
\end{figure}

Further efforts are made to find a universal constraint relation between efficiency and power. Several attempts have been pursued both from the macro-level \cite{constriant,constriant2-1,constriant2}
and the micro-level \cite{trade-off-japan,micro, Holubec2017} with different models.
For low-dissipation heat engine, a simple constraint relation between efficiency $\eta$ and output power $P$ 
\begin{equation}
\widetilde{\eta}+\frac{\left(1-\eta_{\mathrm{C}}\right)\widetilde{P}}{2\left(1+\sqrt{1-\widetilde{P}}\right)-\eta_{\mathrm{C}}\widetilde{P}}\leq1,\label{eq:constraint}
\end{equation}
has been suggested \cite{constriant2-1}, where $\widetilde{\eta}\equiv\eta/\eta_{\mathrm{C}}$ is the normalized
efficiency with the Carnot efficiency and $\widetilde{P}\equiv P/P_{\mathrm{max}}$
is the dimensionless power normalized with the maximum output power $P_{\mathrm{max}}$. It is straightforward to show with Eq.~(\ref{eq:constraint}) that an engine reaches the Carnot bound $\widetilde{\eta}\leq1$ at zero normalized output power $\widetilde{P}=0$, and the efficiency at maximum power is recovered $\widetilde{\eta}\leq1/(2-\eta_{c})$
with $\widetilde{P}=1$, as shown in Fig.~\ref{figure1}. 

Though the analytical derivation of Eq.~(\ref{eq:constraint}) only limited to extreme regions of $\widetilde{P}\simeq0$
and $\widetilde{P}\simeq1$ in Ref.~\cite{constriant2-1}, the constraint Eq.~(\ref{eq:constraint}) works well for all the $\widetilde{P}$, which is checked numerically in the same reference. In this work, we give a succinct analytical derivation of this constraint in the whole region $0\leq\widetilde{P}\leq1$. Furthermore, we obtain a detailed constraint relation Eq.~(\ref{eq:UB}) which also depends on a dimensionless parameter $\zeta$ representing the imbalance between the coupling strengths to the cold and hot heat baths. This detailed constraint relation can provide more information than Eq.~(\ref{eq:constraint}) about how the heat engine parameters affect the upper bound of efficiency at specific output power. In the derivation, we keep temperatures of both hot and cold baths, and cycle endpoints fixed while changing
only operation time.

To validate our results, we present the exact efficiency and output
power of a Carnot-like heat engine with a simple two-level atom as
working substance. Each point in Fig.~\ref{figure1} shows a particular
heat engine cycle with different operation time. In this example,
the evolution of engine is exactly calculated via master equation,
which will be shown in the later discussion. Our model connects microscopic
physical parameters in the cycle to generic parameters in many previous
investigations. All points follow below our constraint curve.

\section{General derivation}

In a finite-time heat engine cycle, we divide the heat exchange $Q_{x}$
with the high ($x=h$) and low ($x=c$) temperature baths into reversible
$Q_{x}^{(r)}=T_{x}\Delta S_{x}$ and irreversible $Q_{x}^{(i)}=-T_{x}\Delta S_{x}^{(i)}$
parts, namely, $Q_{x}=Q_{x}^{(r)}+Q_{x}^{(i)}$, where $\Delta S_{x}^{(i)}$
is the irreversible entropy generated. For the reversible part, we
have $\Delta S_{c}=-\Delta S_{h}$. The low-dissipation assumption \cite{EMP5, LD1, Sasa1997, TuZC2013, LD2, LD3, key-1-esposito}, has been widely used in many
recent studies of finite-time cycle engines, namely 
\begin{equation}
T_{x}\Delta S_{x}^{(i)}=\frac{M_{x}}{t_{x}},\label{eq:M}
\end{equation}
where $t_{x}$ is the corresponding operation time. $M_{x}$ is determined
by the temperature $T_{x}$, the coupling constant to the bath, and
the cycle endpoints, however, not dependent on operation time $t_{x}$.
We will show clearly its dependence on microscopic parameters in the
follow example of two-level atom. The power and efficiency are obtained
simply as $P=(Q_{h}+Q_{c})/(t_{h}+t_{c})$ and $\eta=W/Q_{h}$, where
$W=Q_{h}+Q_{c}$ is the converted work. They can be further expressed
via Eq.~(\ref{eq:M}) and the fact $Q_{h}^{(r)}+Q_{c}^{(r)}=\eta_{C}Q_{h}^{(r)}$
as 
\begin{eqnarray}
P & = & \frac{\eta_{C}Q_{h}^{(r)}-\frac{M_{h}}{t_{h}}-\frac{M_{c}}{t_{c}}}{t_{h}+t_{c}},\label{eq:P}\\
\eta & = & \frac{\eta_{C}Q_{h}^{(r)}-\frac{M_{h}}{t_{h}}-\frac{M_{c}}{t_{c}}}{Q_{h}^{(r)}-\frac{M_{h}}{t_{h}}}.\label{eq:eta}
\end{eqnarray}
Applying the inequality $a/x+bx\geq2\sqrt{ab}$ to Eq.~(\ref{eq:P}),
then we obtain a simple relation between $Q_{h}^{(r)}$ and $P$ as

\begin{equation}
\eta_{C}Q_{h}^{(r)}=P\left(t_{h}+t_{c}\right)+\frac{M_{h}}{t_{h}}+\frac{M_{c}}{t_{c}}\geq2\sqrt{MP},\label{eq:ineq1}
\end{equation}
which defines the maximum output power 
\begin{equation}
P_{\mathrm{max}}\equiv\frac{(\eta_{\mathrm{C}}Q_{h}^{(r)})^{2}}{4M},\label{eq:Pmax}
\end{equation}
with $M=(\sqrt{M_{h}}+\sqrt{M_{c}})^{2}$. We remark here the inequality
Eq.~(\ref{eq:ineq1}) becomes equality only when $t_{h(c)}=\sqrt{M_{h(c)}/P_{\mathrm{max}}}$,
which directly leads to the EMP derived in Ref.~\cite{key-1-esposito}.
This inequality results in $P_{\text{max}}$ because
it reduces the right side of the equality to its infimum and all the
operation times $t_{h(c)}$ are eliminated completely. To obtain a
universal constraint on efficiency and power, we should properly loose
this inequality.

We notice the following fact: a convex function $f(x)$ defined on domain $X$ satisfies
\begin{equation}
\lambda f(x_{1})+(1-\lambda)f(x_{2})\geq f(\lambda x_{1}+(1-\lambda)x_{2}),\label{eq:convex}
\end{equation}
$\forall x_{1},x_{2}\in X$ and $\forall\lambda\in[0,1].$ If we choose the convex function as $f(x)=1/x$ and set $x_{1}=t_h/\sqrt{M_h}$, $x_{2}=t_c/\sqrt{M_c}$, and $\lambda=\sqrt{M_{h}/M}$, it is not hard to find
\begin{eqnarray}
\frac{M_{h}}{t_{h}}+\frac{M_{c}}{t_{c}}  \geq  \frac{M}{t_{c}+t_{h}}.\label{eq:T}
\end{eqnarray}
Take Eq.~(\ref{eq:T}) into Eq.~(\ref{eq:P}), we obtain a constraint
on $\tau\equiv t_{h}+t_{c}$ as 
\begin{equation}
P\tau^{2}-\eta_{\mathrm{C}}Q_{h}^{(r)}\tau+\left(\sqrt{M_{h}}+\sqrt{M_{c}}\right)^{2}\leq0.\label{eq:T2}
\end{equation}
Thus, the total operation time $\tau$ is bounded by $\tau_{-}\leq \tau\leq \tau_{+}$,
with 
\begin{equation}
\tau_{\pm}=\frac{\eta_{\mathrm{C}}Q_{h}^{(r)}}{2P}\left(1\pm\sqrt{1-\tilde{P}}\right).\label{eq:tc}
\end{equation}
Here, $\widetilde{P}\equiv P/P_{\mathrm{max}}$ is the dimensionless
power with $P_{\mathrm{max}}$ given in Eq.~(\ref{eq:Pmax}).

In this work, we mainly concern the upper bound of the efficiency
$\tilde{\eta}_{+}$ for a given power $\tilde{P}$ and fixed engine
setup, i.e., fixed $M_{h(c)}$ and $T_{h(c)}$ (the lower bound $\tilde{\eta}_{-}$
is presented in Appendix A). The problem of finding the upper bound now becomes an optimization
problem: 
\begin{equation}
\tilde{\eta}_{+}=\arg\text{max}(\tilde{\eta})\ \text{subject to }\tau\leq \tau_{+}.\label{eq:optimal}
\end{equation}
Because Eq.~(\ref{eq:eta}) is an increasing function of both $t_{h}$ and
$t_{c}$, the upper bound must be achieved under the condition $\tau=\tau_{+}$. Physically, this fact can be understood as the efficiency
increases as the total operation time increases. Therefore, the solution
of this optimization problem is given by the condition of unique solution
of Eq. (\ref{eq:eta}) and $t_{h}+t_{c}=\tau_{+}$. Straightforwardly, a quadratic
equation for $t_{c}$ can be obtained by taking $t_{h}+t_{c}=\tau_{+}$ into Eq.~(\ref{eq:eta}):
\begin{equation}
t_{c}^{2}+\left[\frac{\left(1-\tilde{\eta}\eta_{C}\right)M_{h}-M_{c}}{\left(1-\tilde{\eta}\right)\eta_{\mathrm{C}}Q_{h}^{(r)}}-\tau_{+}\right]t_{c}+\frac{M_{c}\tau_{+}}{\left(1-\tilde{\eta}\right)\eta_{\mathrm{C}}Q_{h}^{(r)}}=0.\label{eq:tc-2}
\end{equation}
The requirement of unique solution of Eq.~(\ref{eq:tc-2}) (the geometrical explanation of this requirement can be found below [Eq. (A4)] of Appendix A) is equivalent to that the discriminant of the above equation is zero. This immediately results inanother quadratic equation for $\tilde{\eta}_{+}$, the solution of which gives the upper
bound of efficiency for given power and is written explicitly as
\begin{widetext}
\begin{eqnarray}
\tilde{\eta}_{+} & = & \frac{\left(1+\sqrt{1-\tilde{P}}\right)^{2}}{\left(1+\sqrt{1-\tilde{P}}\right)^{2}+\left(1-\frac{\left(1+\zeta\right)^{2}}{4}\eta_{\mathrm{C}}\right)\tilde{P}}+\frac{\left(1-\zeta^{2}\right)\tilde{P}\left(1+\sqrt{1-\tilde{P}}\right)}{\left[\left(1+\sqrt{1-\tilde{P}}\right)^{2}+\left(1-\frac{\left(1+\zeta\right)^{2}}{4}\eta_{\mathrm{C}}\right)\tilde{P}\right]^{2}}\nonumber \\
 &  & \times\left[1-\sqrt{\frac{\eta_{\mathrm{C}}}{2}\left(\frac{\left(1+\zeta\right)^{2}}{4}\eta_{\mathrm{C}}-\zeta\right)\left(1-\sqrt{1-\tilde{P}}\right)+1-\eta_{\mathrm{C}}}-\frac{\left(1+\zeta\right)}{4}\eta_{\mathrm{C}}\left(1-\sqrt{1-\tilde{P}}\right)\right].\label{eq:UB}
\end{eqnarray}
\end{widetext}
Here, we define a dimensionless parameter 
\begin{equation}
\zeta=\frac{\sqrt{M_{h}}-\sqrt{M_{c}}}{\sqrt{M_{h}}+\sqrt{M_{c}}}\in\left[-1,1\right],\label{eq:zeta}
\end{equation}
which characterizes the asymmetry of the dissipation with two heat
baths. In the low-dissipation region, $\tilde{\eta}_{+}$ gives the
highest efficiency when the power and the heat engine setup are assigned.
This upper bound is quite tight according to the simulation results (Appendix A). Moreover, in a wide region of $\tilde{P}$ this
bound is attainable with properly chosen $t_{h}$ and $t_{c}$, though
it is not a supremum for all the $\tilde{P}$.

Usually, we cannot know exactly the heat engine parameters, therefore,
it is useful to find a universal upper bound for all the possible
$\zeta$. As a function of $\zeta$, the analytically proof of the
monotonicity of $\tilde{\eta}_{+}$ is tedious. Instead, we numerically
verified that $\tilde{\eta}_{+}$ is an increasing function in the
whole parameter space, see Appendix B. Thus, the overall bound is reached
at $\zeta=1$, i.e. $M_{h}\gg M_{c}$. We note that a formally similar bound was also obtained in minimal nonlinear irreversible heat engine model \cite{constriant2-1, constriant2}. However, the boundgiven in that model is not equivalent to Eq.~(\ref{eq:constraint}) here.
The definition of $P_{\text{max}}$ in that model is different from Eq.~(\ref{eq:Pmax}) and
depends on $t_{h}$ and $t_{c}$, which can be verified by mapping the parameters wherein back
to ones in the low-dissipation model \cite{Okuda2012}. The detailed discussion can be found in Appendix C

Besides the upper bound, our method also leads to the lower bound for efficiency at
arbitrary given output power,  
\begin{equation}
2\widetilde{\eta}+\sqrt{1-\widetilde{P}}\geq1.
\label{eq:LB}
\end{equation}
The curve for lower bound is illustrated as the gray dotted curve in Fig.~\ref{figure1}.
All the simulated data with two-level atom are above this curve. The detailed derivation for the lower bound is also presented in Appendix A.

We want to emphasize here that this lower bound is different from the lower bound in [Eq. (33)] of Ref.~\cite{constriant2-1}.  The latter one describes the minimum value for maximum efficiency at arbitrary power, which can be derived from Eq.~(\ref{eq:UB}) by choosing $\zeta=-1$. Yet, the lower bound we obtained in Eq.~(\ref{eq:LB}) determines the minimum possible efficiency for the  arbitary given value of power.

\begin{figure}
\includegraphics[width=8.5cm]{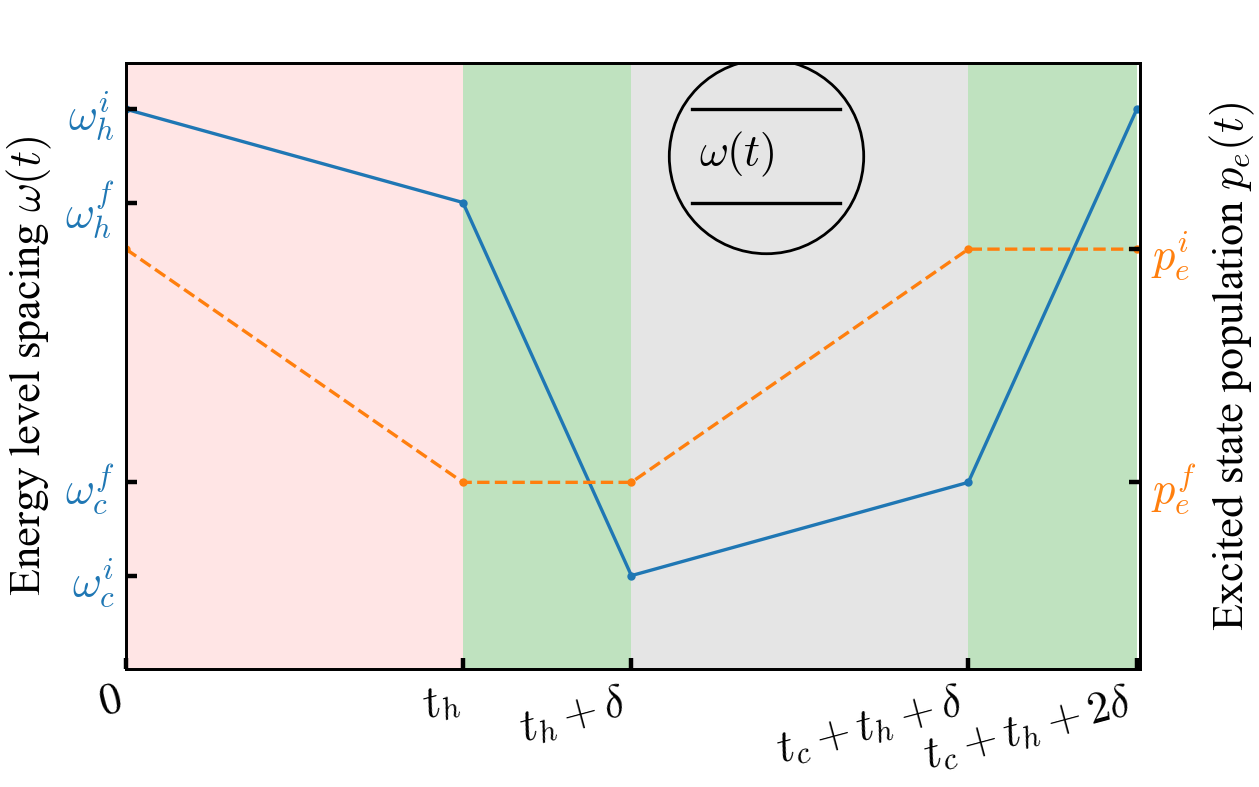}

\caption{\label{fig2}(Color online) Carnot-like cycle with four strokes. (i) $\left[0,t_{h}\right]$
quasi-isothermal process in contact with hot bath. (ii) $[t_{h},t_{h}+\delta]$
adiabatic process. (iii) $\left[t_{h}+\delta,t_{c}+t_{h}+\delta\right]$
quasi-isothermal process. (iv) $\left[t_{c}+t_{h}+\delta,t_{c}+t_{h}+2\delta\right]$
adiabatic process in contact with cold bath. The blue solid curve
shows the change of energy spacing $\omega\left(t\right)$, and the
orange dotted curve show the evolution of excited state population.}
\end{figure}

To achieve the maximum efficiency at given normalized power $\tilde{P}$,
we adjust three parameters: the operation time $t_{h}$ and $t_{c}$
during contacting with both hot and cold baths, and the entropy generation
ratio $\zeta$, while fixing the temperatures $T_{h}$, $T_{c}$,
and the reversible heat exchange $Q_{h}^{(r)}$. The derivation leaves
a question about adjusting $\zeta$, namely tuning $M_{h}$ and $M_{c}$.
In our previous discussion, $M_{h}$ and $M_{c}$ are phenomenological
assumed without connecting to the physical parameters. In our example
of two-level atomic heat engine, tuning $M_{x}$ $(x=h,c)$ is achieved
via changing the coupling constant of heat engine to bathes. We now
switch to a specific Carnot-like quantum heat engine with two-level
atom.

\section{Validate with two-level quantum heat engine}

Quantum heat engine with two-level atom is the simplest engine to
illustrate the relevant physical mechanisms \cite{CPSun2007,TSL}. Here,
we design a Carnot-like cycle with two-level atom, whose energy levels
(the excited state $\left|e\right\rangle $ and ground state $\left|g\right\rangle $)
are tuned by the outsider agent to extract work, namely $H=\frac{1}{2}\omega\left(t\right)\sigma_{z}$,
where $\sigma_{z}=\left|e\right\rangle \left\langle e\right|-\left|g\right\rangle \left\langle g\right|$
is the Pauli matrix in z-direction. The finite-time cycle consists
of four strokes. Operation time per cycle is $\tau=t_{h}+t_{c}+2\delta$,
where $t_{h}$ ($t_{c}$) is the interval of quasi-isothermal process
in contact with hot (cold) bath and $\delta$ is the interval of adiabatic
process. The quasi-isothermal process retains to the normal isothermal
process at the limit $t_{h(c)}\rightarrow\infty$. The cycle is illustrated
with Fig.~\ref{fig2}:

(i) Quasi-isothermal process in contact with hot bath ($0<(t\;\mathrm{mod}\;\tau)<t_{h}$):
The energy spacing change is linearly decreases as $\omega\left(t\right)=\omega_{h}^{i}+v_{h}t$,
where $v_{h}=\epsilon_{h}/t_{h}$ is the changing speed with both $\omega_{h}^{i}$
and $\omega_{h}^{f}=\omega_{h}^{i}+\epsilon_{h}$ fixed. The change
of energy spacing is shown as solid-blue curve in Fig.~\ref{fig2}.
The linear change of the energy spacing is one of the simplest protocols.

(ii) Adiabatic process ($t_{h}<(t\;\mathrm{mod}\;\tau)<t_{h}+\delta$):
The energy level spacing is further reduced from $\omega_{h}^{f}$
to $\omega_{c}^{i}$, while it is isolated from any heat bath. Since
there is no transition between the two energy levels, the interval
$\delta$ of the adiabatic process is irrelevant of the thermodynamical
quantities. In the simulation, we simply use $\delta=0$. The heat
exchange is zero, and the entropy of the system remains unchanged.

(iii) Quasi-isothermal process in contact with cold bath ($t_{h}+\delta<(t\;\mathrm{mod}\;\tau)<t_{h}+\delta+t_{c}$).
The process is similar to the first process, yet the energy spacing
$\omega\left(t\right)=\omega_{c}^{i}+v_{c}\left(t-t_{h}-\delta\right)$
increases with speed $v_{c}=\epsilon_{c}/t_{c}$ and ends at $\omega_{c}^{f}=\omega_{c}^{i}+\epsilon_{c}$.

(iv) Adiabatic process ($t_{h}+\delta+t_{c}<(t\;\mathrm{mod}\;\tau)<t_{h}+2\delta+t_{c}$).
The energy spacing is recovered to the initial value $\omega_{h}^{i}$.

The two-level atom operates cyclically following the above four strokes,
whose dynamics is described by the master equation

\begin{equation}
\frac{dp_{e}\left(t\right)}{dt}=-\kappa\left(t\right)p_{e}\left(t\right)+C\left(t\right),\label{eq:master equation}
\end{equation}
where $p_{e}\left(t\right)\equiv\langle e\vert\hat{\rho}\left(t\right)\vert e\rangle$
is the excited state population of the density matrix $\hat{\rho}\left(t\right)$,
$C\left(t\right)=\gamma\left(t\right)n[\omega(t)]$ and $\kappa\left(t\right)=\gamma\left(t\right)\left(2n[\omega(t)]+1\right)$
with $n[\omega(t)]=1/\left(\exp[\beta\left(t\right)\omega\left(t\right)]-1\right)$
is the mean occupation number of bath mode with frequency $\omega\left(t\right)$.
The dissipative rate $\gamma\left(t\right)$ is a piecewise function
which is a constant $\gamma_{h}$ ($\gamma_{c}$) during quasi-isothermal
processes (i) and (iii), and zero during the two adiabatic processes.
The inverse temperature $\beta\left(t\right)$ is also a piecewise
function defined on quasi-isothermal processes (i) and (iii) with values
$\beta_{h}$ and $\beta_{c}$, respectively. In this work we assume the energy levels always avoid crossing during the whole cycle, thus the quantum adiabatic theorem guarantees the master equation does not involve the contribution of coherence induced by non-adiabatic transition \cite{Zanardi2012,Ogawa2017,Kosloff2018}. In other words, the two-level quantum heat engine we study here is working in the classical regime. 

\begin{figure}
\includegraphics[width=8.5cm]{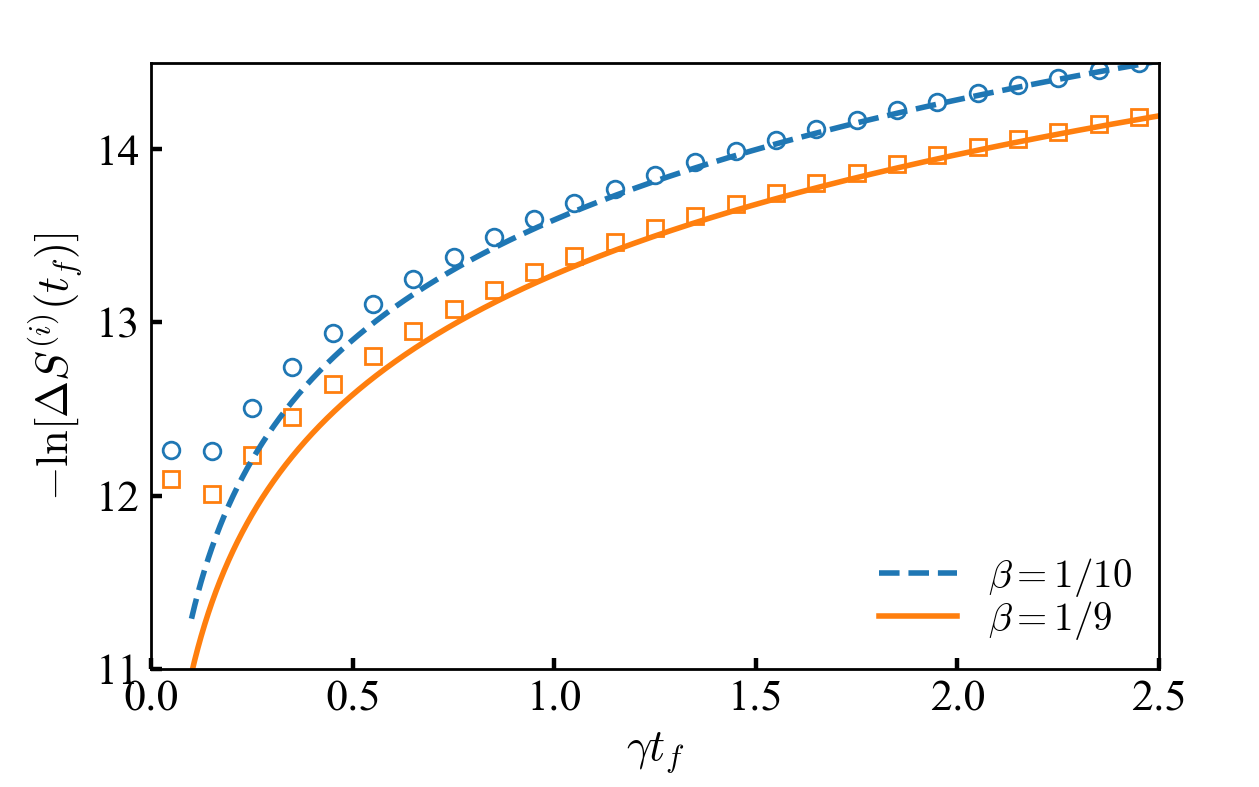}

\caption{\label{fig3}(Color online) Irreversible entropy generation as a function of operation time at the temperature of the hot bath $\beta=1/10$ (orange) and the cold bath $\beta=1/9$ (blue). The points show the exact numerical results,
and lines show the analytical results of the high temperature limit Eq.~(\ref{eq:irriverisbleentropy-1}).}
\end{figure}

In the simulation, we have chosen an arbitrary initial state, and
perform the calculation of both efficiency and output power after
the engine reaches a steady state, in which the final state of stroke
(iv) matches the initial state of stroke (i). Different from the textbook
Carnot cycle with isothermal process, the microscopic heat engine operates
away from equilibrium in the finite-time Carnot-like cycle with the
quasi-isothermal process. For infinite operation time $\left(t_{h},t_{c}\right)$,
the current cycle recovers the normal Carnot cycle.

To get efficiency and power, we consider the heat exchange and work
done in two quasi-isothermal processes. The internal energy change
and work done in stroke (i) is $\Delta U_{h}=\mathrm{Tr}[H(t_{h})\tilde{\rho}(t_{h})-H(0)\tilde{\rho}(0)]$
and $W_{h}=\mathrm{Tr}[\int_{0}^{t_{h}}\frac{dH(t)}{dt}\tilde{\rho}(t)dt]$,
respectively. The total heat absorbed from the hot bath is given via
the first law of thermodynamics as $Q_{h}=\Delta U_{h}+W_{h}$. The
same calculation can be carried out for $Q_{c}$ in stroke (iii) with
the initial and final times are substituted by $t_{h}+\delta$ and
$t_{h}+\delta+t_{c}$. The work converted and the efficiency are defined
the same as in the general discussion. In our simulation, we have
fixed energy spacing of the two-level atom at the four endpoints:
$\omega_{h}^{i}$, $\omega_{h}^{f}$, $\omega_{c}^{i}$, $\omega_{c}^{f}$.

To check the upper bound, we have generated the efficiency and output
power with different operation times. Each point in Fig.~\ref{figure1}
corresponds to a set of different operation time $\left(t_{h},t_{c}\right)$.
In all the simulations, the operation time $t_{h}$ and $t_{c}$ are
randomly generated. All points fall perfectly under the upper bound
shown in Eq.~(\ref{eq:constraint}).

To be comparable with the general analysis above, it is meaningful
to check two key conditions: (1) low-dissipation region with $1/t$
scaling of irreversible entropy production, and (2) the value of tuning
parameters $\zeta$. To check the two conditions, we firstly need
calculate the irreversible entropy generation. Here, we consider a
generic quasi-isothermal process starts at $t=0$ and end at $t=t_{f}$
with $\omega\left(t\right)=\omega_{0}+\epsilon t/t_{f}$. To simplify
the discussion, we remove the index $h$ and $c$ related to the bath.
The solution to Eq.~(\ref{eq:master equation}) is formally obtained
as $p_{e}\left(t\right)=\textrm{e}^{-\int_{0}^{t}\kappa\left(t_{1}\right)dt_{1}}[p_{e}\left(0\right)+\int_{0}^{t}\textrm{e}^{\int_{0}^{t_{1}}\kappa\left(t_{2}\right)dt_{2}}C\left(t_{1}\right)dt_{1}]$,
$t\in[0,t_{f}]$. The entropy change during the process is evaluated
via von Neuman formula $S(\hat{\rho})=-k_{B}\mathrm{Tr}[\hat{\rho}\ln\hat{\rho}]$
as $\Delta S(t_{f})=S(\hat{\rho}(t_{f}))-S(\hat{\rho}(0))$. The irreversible
entropy production in this quasi-isothermal process reads $\Delta S^{(i)}=\Delta S\left(t_{f}\right)-\beta Q$,
where exchange $Q$ is obtained via $Q=\Delta U+W$.

At the high temperature limit $\beta\omega\left(t\right)\ll1$, and
for $\omega_{0}\gg\left|\epsilon\right|$, namely, the linear response
region, the irreversible entropy production reads
$\Delta S^{(i)}\approx\frac{\left(\beta\epsilon\right)^{2}}{4\tilde{\gamma}t_{f}}\left(1-\frac{1-e^{-\tilde{\gamma}t_{f}}}{\tilde{\gamma}t_{f}}\right)$,
where $\widetilde{\gamma}\equiv2\gamma/\left(\beta\omega_{0}\right)$ (see Appendix D).
At long-time limit $\tilde{\gamma}t_{f}\gg1$, we keep only the leading
term and get the normal assumption of $1/t$ behavior of entropy generation
\begin{equation}
\Delta S^{(i)}\approx\frac{\left(\beta\epsilon\right)^{2}}{4\tilde{\gamma}t_{f}}.\label{eq:irriverisbleentropy-1}
\end{equation}
A general discussion about the $1/t$ form of the irreversible entropy generation based on stochastic thermodynamics can be found in Ref.~\cite{Sasa1997}. We plot the irreversible entropy generation as a function of contacting
time $t_{f}$ in Fig.~\ref{fig3}. The points show the exact entropy
generation by solving Eq.~(\ref{eq:master equation}). At short time
$\tilde{\gamma}t_{f}<1$, the entropy deviates from the low-dissipation
region. Especially, in the extremely short contact time limit, $\lim_{t_{f}\rightarrow0}\Delta S^{(i)}=\left(\beta\epsilon\right)^{2}/8$
is a finite quantity instead of be divergent as in $1/t$ assumption.
To reach this low-dissipation limit, we need either large coupling
$\gamma$ between system and bath, or long-time contacting time $t_{f}>1/\tilde{\gamma}$.
In the simulation, we have chosen the operation time $t_{h}$ and
$t_{c}$ to fulfill this requirement.

Back to the example of two-level atomic Carnot-like heat engine, the
parameter $M_{x}$ ($x=h,c$) is simply $M_{x}\equiv\beta_{x}^{2}\omega_{x}^{i}\epsilon_{x}^{2}/(8\gamma_{x})$,
and $\gamma_{x}$ is the only parameter available to the tune $M_{x}$.
Therefore, the dimensionless parameter $\zeta$ for whole cycle can
be tuned via $\gamma_{h}$ and $\gamma_{c}$. In the simulation in Fig.~\ref{figure1}, we
have the parameters $\eta_C=0.1$ and $\zeta=0.5$. In this region the upper bound is very close the one with $\zeta=1$.

We remark that the current proof of the upper bound is based on assumption
of low-dissipation. Taking the two-level atomic example, this assumption together with the microscopic expression for $M_{x}$
is guaranteed in the long time limit $\gamma t_{f}\gg\beta|\epsilon|$ and with the requirement $\omega\gg|\epsilon|$.
It is interesting to note that low-dissipation can be achieved with
large coupling strength $\gamma_{x}$, according to Eq.~(\ref{eq:irriverisbleentropy-1}).
However, it remains open to obtain the universal bound for system
beyond low-dissipation region, which will be discussed elsewhere.

\section{Conclusion}

In summary, we have derived the constraint relation between efficiency
and output power of heat engine working under the so-called low-dissipation
region. A general proof of the constraint for all the region of output power
is given. We also obtain a detailed constraint depending on the dissipation to the hot and cold baths, which can provide more information for a specific heat engine model. Moreover, in a concrete example
of heat engine with two-level atom, we connect phenomenological
parameters to the microscopic parameters, such as coupling constants to baths. These connections enable practical
adjusting the heat engine to achieve the designed function via optimizing
the physical parameters, and can be experimentally verified with the
state of art superconducting circuit system~\cite{Pekola2009}. 
\begin{acknowledgments}
Y.~H.~Ma is grateful to Z.~C.~Tu, S.~H.~Su, and J.~F.~Chen for the
helpful discussion. This work is supported by NSFC (Grants No.~11534002), the National Basic Research Program
of China (Grant No.~2016YFA0301201 \& No.~2014CB921403), the NSAF
(Grant No.~U1730449 \& No.~U1530401) and Beijing Institute of Technology
Research Fund Program for Young Scholars. 
\end{acknowledgments}

\appendix
\section{Lower bound of efficiency}

The lower bound is obtained via the constraint $t_{h}+t_{c}=\tau_{-}$, take this equation into Eq.~(3) of the main text, we have an equation of $t_{h}$ 
\begin{equation}
t_{h}^{2}-\left(\tau_{-}+\frac{M_{h}-M_{c}}{\eta_{\mathrm{C}}Q_{h}^{(r)}-P\tau_{-}}\right)t_{h}+\frac{M_{h}\tau_{-}}{\eta_{\mathrm{C}}Q_{h}^{(r)}-P\tau_{-}}=0.\label{eq:th}
\end{equation}
This equation has only one solution 
\begin{equation}
t_{h}=\frac{2\left(\sqrt{M_{h}}+\sqrt{M_{c}}\right)^{2}\left(1+\zeta\right)}{\eta_{\mathrm{C}}Q_{h}^{(r)}\left(1+\sqrt{1-\tilde{P}}\right)},\label{eq:th2}
\end{equation}
This solution together with Eq.~(4) of the main text gives the lower bound
of efficiency 
\begin{equation}
\tilde{\eta}_{-}=\frac{1}{2}\frac{1-\sqrt{1-\tilde{P}}}{1-\frac{1}{8}\eta_{\mathrm{C}}\left(1+\zeta\right)\left(1+\sqrt{1-\tilde{P}}\right)}.\label{eq:LB}
\end{equation}
This lower bound gives the information of the worst efficiency for
a given low-dissipation heat engine at power $\tilde{P}$. Similar
to $\tilde{\eta}_{+}$, $\tilde{\eta}_{-}$ is not the infimum for
all the $\tilde{P}$ either.

As $\tilde{\eta}_{-}$ is obvious a decreasing function of $\zeta$,
the universal lower bound is at $\zeta=-1$, thus we have 
\begin{equation}
\tilde{\eta}\geq\frac{1}{2}\left(1-\sqrt{1-\tilde{P}}\right).\label{eq:uLB}
\end{equation}

\begin{figure}[h]
\includegraphics[width=8.5cm]{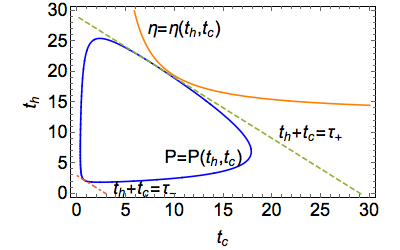} \caption{\label{fig:geo}(Color online) The curves $P=P(t_{h},t_{c})$ and $\eta=\eta(t_{h},t_{c})$
are plotted in closed blue line and open orange line, respectively; $t_{h}+t_{c}=\tau_{+}$
and $t_{h}+t_{c}=\tau_{-}$ are plotted with greed dashed line and red
dot-dashed line, respectively. In this example, we choose $M_{h}=9$, $M_{c}=1$,
$Q_{h}^{(r)}=10$, $\eta_{C}=0.6$ and $P=0.6P_{\text{max}}$, $\eta=0.95\eta_{C}$.}
\end{figure}

Notice that the way we solve the lower bound is a little different
from that of the upper bound. For lower bound, we directly solve the
constraint $t_{h}+t_{c}=\tau_{-}$ with Eq.~(3), instead of Eq.~(4) of the main text.
This can be well understood by plotting $P=P(t_{h},t_{c})$ [Eq.~(3) of the main text]
and $\eta=\eta(t_{h},t_{c})$ [Eq.~(4) of the main text] on the plane
spanned by $t_{h}$ and $t_{c}$, as shown in Fig.~\ref{fig:geo}.
In the first quadrant, $P=P(t_{h},t_{c})$ appears as the blue closed
curve and $\eta=\eta(t_{h},t_{c})$ as the orange open curve. The
intersections of $P=P(t_{h},t_{c})$ and $\eta=\eta(t_{h},t_{c})$
gives the physically attainable $t_{h}$ and $t_{c}$ for given $P$
and $\eta$. Two tangent lines $t_{h}+t_{c}=\tau_{+}$ (green dashed
line) and $t_{h}+t_{c}=\tau_{-}$ (red dot-dashed line) sandwich $P=P(t_{h},t_{c})$
in between. As $\eta$ is an increasing function of both $t_{h}$
and $t_{c}$, the larger $\eta$ is, the curve $\eta(t_{h},t_{c})$
is more away from the origin of coordinates, and vice versa. With
this fact, it is not hard to see, all the curves $\eta(t_{h},t_{c})$
on the right side of $t_{h}+t_{c}=\tau_{+}$ have $\eta$ larger than
any possible $\eta$ with $P$ given. Therefore, the curve $\eta=\eta(t_{h},t_{c})$
which is tangent with $t_{h}+t_{c}=\tau_{+}$ gives the least upper bound
we can find. On the other hand, the curve $P=P(t_{h},t_{c})$ itself
is already on the right side for tangent line $t_{h}+t_{c}=\tau_{-}$,
thus the tangent point leads to the largest lower bound we can find.

\begin{figure}[h]
\includegraphics[width=8cm]{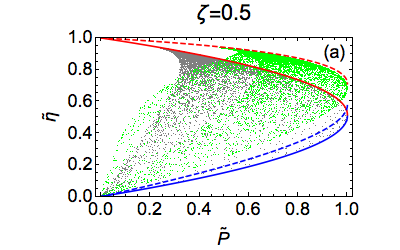} 
\includegraphics[width=8cm]{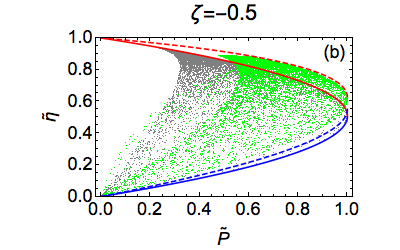}

\caption{\label{fig:simu}(Color online) The upper (lower) bound with $\eta_{C}=0.1$ and
$\eta_{C}=0.9$ are plotted with red (blue) solid and red (blue) dashed
line, respectively, with (a) $\zeta=0.5$ and (b) $\zeta=-0.5$. The
gray dots and green dots are plotted with random $t_{h}$ and $t_{c}$.}
\end{figure}

To show how close the upper and lower bounds to the attainable $\tilde{\eta}(\tilde{P}),$
we plot these two bounds with randomly simulated points $(\tilde{P},\tilde{\eta})$
in Fig.~\ref{fig:simu}. The upper and lower bounds are calculated
by Eq.~(13) of the main text and Eq.~(\ref{eq:LB}), respectively, and
the simulation points are spotted according to Eqs.~(3,
4) of the main text with randomly chosen $t_{h}$ and $t_{c}$. We can
see these two bounds are quite tight that the simulated points are
nearly saturate with them.

\section{The monotonicity of $\widetilde{\eta}_{+}$}

\begin{figure}
\includegraphics[width=8cm]{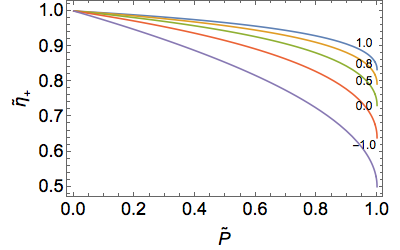} 
\includegraphics[width=8cm]{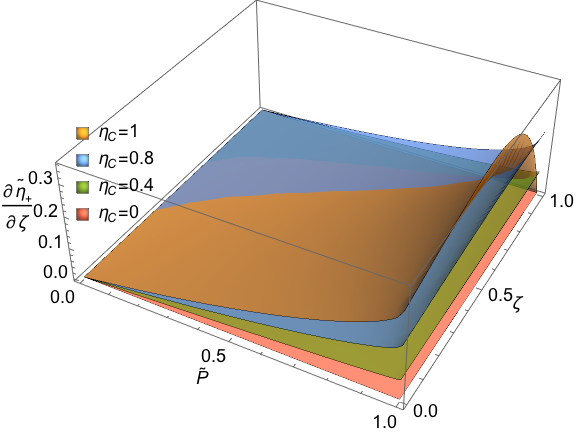}
\caption{\label{fig:upperbounds}(Color online) Upper panel: The upper bound $\tilde{\eta}_{+}$ as a function
of $\tilde{P}$, the curves from bottom to up is in the order of increasing
$\zeta$ with $\zeta=-1.0,0.0,0.5,0.8,1.0$. Here we choose $\eta_{C}=0.8$. Lower panel: The derivative of $\partial\tilde{\eta}_{+}/\partial\zeta$ with respective to $\tilde{P}$ and $\zeta$. From above to bottom,
the surfaces are plotted with decreasing $\eta_{C}=1.0,0.8,0.4,0.0$.
It can be seen the derivative is non-negative in the whole parameter
space, thus proves $\tilde{\eta}_{+}\left(\zeta\right)$ is an increasing
function.}
\end{figure}

The upper bound of efficiency $\widetilde{\eta}_{+}$ is an increasing
function of $\zeta$. As illustrated in the left panel of Fig.~\ref{fig:upperbounds},
the curves of $\widetilde{\eta}_{+}(\zeta)$ are in the order of increasing
$\zeta$ from bottom to up. If $\eta_{C}$ is getting smaller, the
difference of $\widetilde{\eta}_{+}(\zeta)$ between different $\zeta$
disappears gradually. 

As the expression of $\tilde{\eta}_{+}$ is complicated, the analytical
proof of its monotonicity is tedious and difficult. Instead numerical
calculate the derivative of $\tilde{\eta}_{+}(\zeta)$. As we can
see in the counter plot in the right panel of Fig.~\ref{fig:upperbounds}, $\partial\tilde{\eta}_{+}/\partial\zeta$
is non-negative in all the parameter region of $\tilde{P}$ and $\zeta$,
thus $\tilde{\eta}_{+}$ is indeed an increasing function of $\zeta$
and its maximum value is at $\zeta=1$.

\section{Compare with the minimally nonlinear irreversible heat engine model}

Based on then extended Onsager relations, a model named as ``minimally nonlinear irreversible heat engine'' was proposed to study the same problem about the relation between efficiency and power. It is usually believed that the minimally nonlinear irreversible heat engine and the low-dissipation heat engine model, since there is a one-to-one mapping between the parameters of the two models \cite{Okuda2012}. Recently, a formally same constraint as Eq.~(\ref{eq:constraint}) in the main text is obtained by the nonlinear irreversible heat engine model \cite{constriant2}. However, we have to emphasise that, even though there is the equivalence of the two models and the similar results they give, the bounds on efficiency at arbitrary power given by them are different. The reason can be ascribed to the optimization parameters in the two models are essentially different.

Specifically speaking, the  definition of the max power $P_{\text{max}}$
in Ref.~\cite{constriant2} is different from the one in this work. This can be verified by mapping
the $P_{\text{max}}$ of Eq.~(11) in Ref.~\cite{constriant2} from the minimally nonlinear
irreversible model back to the low-dissipation model. If we express the
$P_{\text{max}}$ in Ref.~\cite{constriant2} with the parameters in the low-dissipation
model, it actually depends on $t_{h}$ and $t_{c}$. Explicitly, in Ref.~\cite{constriant2} $P_{\text{max}}$ is defined as
\begin{equation}
P_{\text{max}}=\frac{q^{2}L_{22}\eta_{C}^{2}}{4T_{c}},\label{eq:Pmax19}
\end{equation}
where $L_{22}$ is one of the Onsager coefficients.
The mapping of the parameters of the two heat engine models is given in Ref.~\cite{Okuda2012} as,
\begin{eqnarray}
L_{22} & =\frac{T_{h}^{2}T_{c}\Delta S^{2}}{(T_{h}\Sigma_{h}+T_{c}\Sigma_{c}/\alpha)(\alpha+1)},\\
\alpha & =t_{c}/t_{h}.
\end{eqnarray}
In the tight-coupling case ($q=1$), and with the notation $M_{h(c)}=T_{h(c)}\Sigma_{h(c)}$, we can see the Eq.~(\ref{eq:Pmax19}) reads
\begin{eqnarray}
P_{\text{max}} & =\frac{T_{h}^{2}\Delta S^{2}}{(T_{h}\Sigma_{h}+T_{c}\Sigma_{c}/\alpha)(\alpha+1)}\frac{\eta_{C}^{2}}{4}\nonumber \\
 & =\frac{T_{h}^{2}\Delta S^{2}}{(\frac{M_{h}}{t_{h}}+\frac{M_{c}}{t_{c}})(t_{c}+t_{h})}\frac{\eta_{C}^{2}}{4}\nonumber \\
 & =\frac{(\eta_{\mathrm{C}}Q_{h}^{(r)})^{2}}{4(\frac{M_{h}}{t_{h}}+\frac{M_{c}}{t_{c}})(t_{c}+t_{h})},\label{eq:P19}
\end{eqnarray}
which is obviously different from the max power in Eq.~(6) in the main text. Therefore, Eq.~(1) in this work and Eq.~(22) in Ref.~\cite{constriant2} are intrinsically different.\\

It can be seen from Eq.~(\ref{eq:P19}) that $P_{\text{max}}$ still depends on $t_{h}$ and $t_{c}$ in Ref.~\cite{constriant2}, thus another step of optimization with respect to $\alpha$ is needed to arrive at the real max power, this fact is already indicated in the Ref.~\cite{Okuda2012}. We would like to emphases here, the equivalence
of the two models only means there exists a mapping between parameters
of these two models, which does not imply the optimization processes and the bounds are the same.

\begin{widetext}
\section{Irreversible entropy generation}
In this section, we show detailed derivation of irreversible entropy
generation of a TLA in a quasi-isothermal process. Here, we focus
on the case the energy gap of the TLA is linearly changed, this minimal
model is enough to illustrate the validity and limitations of the
low-dissipation assumption. The more general time-dependent cases
will be discussed elsewhere. The Born-Markov master equation Eq.~(\ref{eq:master equation}) of the main text
in the main text is capable of the case that the Hamiltonian has no
level crossing. It can be formally solved as 
\begin{eqnarray}
p_{e}\left(t\right) & = & e^{-\int_{0}^{t}\kappa\left(t_{1}\right)dt_{1}}\left[p_{e}\left(0\right)+\int_{0}^{t}e^{\int_{0}^{t_{1}}\kappa\left(t_{2}\right)dt_{2}}C(t_{1})dt_{1}\right],
\end{eqnarray}
where 
\begin{equation}
C\left(t\right)=\frac{\gamma}{\exp[\beta\omega\left(t\right)]-1},\ \mathrm{and}\ \kappa\left(t\right)=\gamma\coth\left[\frac{\beta\omega\left(t\right)}{2}\right].
\end{equation}
Integrated by parts, we have 
\begin{eqnarray}
p_{e}\left(t\right) & = & p_{e}^{0}\left(t\right)+\frac{\beta\epsilon}{4t_{f}}\int_{0}^{t}\frac{e^{-\int_{t_{1}}^{t}\kappa\left(t_{2}\right)dt_{2}}}{\cosh^{2}\left[\frac{\beta\omega(t_{1})}{2}\right]}dt_{1}+\left[p_{e}\left(0\right)-p_{e}^{0}\left(0\right)\right]e^{-\int_{0}^{t}\kappa\left(t_{1}\right)dt_{1}}.\label{eq:pe}
\end{eqnarray}
Here we define 
\[
p_{e}^{0}\left(t\right)\equiv\frac{1}{\exp[\beta\omega(t)]+1},
\]
which is the excited state population when the TLA is equilibrium
with the heat bath. For the sake of simplification, we assume the
initial state of the finite-time isothermal process is an equilibrium
state, thus the last term of Eq.~(\ref{eq:pe}) can be ignored. Now
we can discuss the high and low temperature behavior of the irreversible
entropy production in such process.

\subsection{High temperature limit}

In the high temperature limit, $\beta\omega_{0}\ll1$, the integration
in Eq.~(\ref{eq:pe}) is approximated as 
\[
\int_{0}^{t}\kappa\left(t_{1}\right)dt_{1}=\gamma\left(t+\frac{2t_{f}}{\beta\epsilon}\ln\frac{1-e^{-\beta\omega(t)}}{1-e^{-\beta\omega_{0}}}\right)\approx\gamma\left[t+\frac{2t_{f}}{\beta\epsilon}\ln\left(1+\frac{\epsilon t}{\omega_{0}t_{f}}\right)\right],
\]
which can be further written as, with the assumption $\omega_{0}\gg\left|\epsilon\right|$
\begin{eqnarray}
\int_{0}^{t}\kappa\left(t_{1}\right)dt_{1} & \approx & \gamma t\left(1+\frac{2}{\beta\omega_{0}}\right)\approx\frac{2\gamma t}{\beta\omega_{0}}\equiv\tilde{\gamma}t.\label{eq:gammat}
\end{eqnarray}
Here we define an effective dissipative rate $\tilde{\gamma}$ which
is in inverse of $\beta$ and $\omega_{0}$. Therefore, in the high
temperature limit the excited population $p_{e}\left(t\right)$ reads

\begin{equation}
p_{e}\left(t\right)\approx p_{e}^{0}\left(t\right)+\frac{\beta\epsilon}{4\tilde{\gamma}t_{f}}\left(1-e^{-\tilde{\gamma}t}\right).\label{eq:pe1}
\end{equation}

Next, the irreversible entropy production is given straightforwardly
by definition, 
\begin{eqnarray}
\Delta S^{(i)} & = & \Delta S\left(t_{f}\right)-\beta Q.\label{eq:IS}
\end{eqnarray}
The heat absorbed from the bath is given by 
\begin{eqnarray}
Q & = & \Delta U-W\nonumber \\
 & = & \omega(t_{f})p_{e}(t_{f})-\omega_{0}p_{e}\left(0\right)-\frac{1}{\beta}\ln\frac{1+e^{-\beta\omega_{0}}}{1+e^{-\beta\omega(t_{f})}}-\frac{\beta\epsilon^{2}}{4\tilde{\gamma}t_{f}}\left(1-\frac{1-e^{-\tilde{\gamma}t_{f}}}{\tilde{\gamma}t_{f}}\right),\label{eq:Q-hT}
\end{eqnarray}
and the entropy change of the system reads 
\begin{eqnarray}
\Delta S(t_{f}) & = & -\text{Tr}\left[\hat{\rho}(t_{f})\ln\hat{\rho}(t_{f})\right]+\text{Tr}\left[\hat{\rho}(0)\ln\hat{\rho}(0)\right]\nonumber \\
 & = & \beta\left[\omega(t_{f})p_{e}(t_{f})-\omega_{0}p_{e}(0)-\frac{1}{\beta}\ln\frac{1+e^{-\beta\omega_{0}}}{1+e^{-\beta\omega(t_{f})}}\right]\nonumber \\
 &  & -p_{e}(t_{f})\ln\left[1+\frac{\beta\epsilon}{4p_{e}^{0}(t_{f})\tilde{\gamma}t_{f}}\left(1-e^{-\tilde{\gamma}t_{f}}\right)\right]-p_{g}(t_{f})\ln\left[1-\frac{\beta\epsilon}{4p_{g}^{0}(t_{f})\tilde{\gamma}t_{f}}\left(1-e^{-\tilde{\gamma}t_{f}}\right)\right].\label{eq:S-hT}
\end{eqnarray}
The first term of Eq.~(\ref{eq:S-hT}) is the entropy change in a
quasi-static isothermal process with the same initial and final energy
spacings, which can be canceled with the first three terms of Eq.~(\ref{eq:Q-hT}).
The last two terms of Eq.~(\ref{eq:S-hT}) are related to the entropy
difference between the real finial state $p_{e/g}(t_{f})$ and the
equilibrium state $p_{e/g}^{0}(t_{f})$, the leading term of which
is of the order of $t_{f}^{-2}$ in the high temperature limit: 
\begin{eqnarray*}
 &  & -p_{e}(t_{f})\ln\left[1+\frac{\beta\epsilon}{4p_{e}^{0}(t_{f})\tilde{\gamma}t_{f}}\left(1-e^{-\tilde{\gamma}t_{f}}\right)\right]-p_{g}(t_{f})\ln\left[1-\frac{\beta\epsilon}{4p_{g}^{0}(t_{f})\tilde{\gamma}t_{f}}\left(1-e^{-\tilde{\gamma}t_{f}}\right)\right]\\
 & \approx & \left[\frac{p_{g}(t_{f})}{p_{g}^{0}(t_{f})}-\frac{p_{e}(t_{f})}{p_{e}^{0}(t_{f})}\right]\frac{\beta\epsilon}{4\tilde{\gamma}t_{f}}\left(1-e^{-\tilde{\gamma}t_{f}}\right)=-\frac{1}{p_{g}^{0}(t_{f})p_{e}^{0}(t_{f})}\left[\frac{\beta\epsilon}{4\tilde{\gamma}t_{f}}\left(1-e^{-\tilde{\gamma}t_{f}}\right)\right]^{2}.
\end{eqnarray*}

Therefore, by substituting Eqs.~(\ref{eq:Q-hT}) and (\ref{eq:S-hT})
into Eq.~(\ref{eq:IS}), the irreversible entropy production then reads

\begin{eqnarray}
\Delta S^{(i)} & \approx & \frac{\beta^{2}\epsilon^{2}}{4\tilde{\gamma}t_{f}}\left[1-\frac{1-e^{-\tilde{\gamma}t_{f}}}{\tilde{\gamma}t_{f}}\left(1+\frac{1-e^{-\tilde{\gamma}t_{f}}}{4p_{g}^{0}(t_{f})p_{e}^{0}(t_{f})}\right)\right].\label{eq:Si-HT}
\end{eqnarray}
When $\tilde{\gamma}t_{f}\gg1$, we keep only the leading term, thus
we have 
\begin{equation}
\Delta S^{(i)}\left(t_{f}\gg\tilde{\gamma}^{-1}\right)=\frac{\left(\beta\epsilon\right)^{2}}{4\tilde{\gamma}t_{f}},
\end{equation}
which is the result presented in the main text. In this minimal model,
the low-dissipation assumption is valid when the operation time is
longer than the time scale of $\tilde{\gamma}^{-1}$. When $\tilde{\gamma}t_{f}\ll1$,
the irreversible entropy production has a finite limitation: 
\begin{equation}
\Delta S^{(i)}\left(t_{f}\ll\tilde{\gamma}^{-1}\right)=\frac{\left(\beta\epsilon\right)^{2}}{8}.
\end{equation}
In this short time region, the low-dissipation assumption is not satisfied
anymore, thus the constrain relation between efficiency and power
discussed in the main text is not applicable in this case.

\subsection{Low temperature limit}

We can also obtain an approximated analytical result of the irreversible
entropy production in the low temperature region $\beta\omega_{0}\gg1$.
By using the fact $\kappa\left(t\right)\approx\gamma$ and $\cosh^{2}\left[\frac{\beta\omega(t_{4})}{2}\right]\approx\exp[\beta\omega(t)]/4$ for low temperature, the excited state population can be approximated as

\begin{eqnarray}
p_{e}\left(t\right) & \approx & p_{e}^{0}(t)+\frac{\beta\epsilon}{t_{f}}\int_{0}^{t}e^{-\gamma(t-t_{1})-\beta\omega(t_{1})}dt_{1}\nonumber \\
 & = & p_{e}^{0}(t)+\frac{\beta\epsilon}{\gamma t_{f}-\beta\epsilon}\left[e^{-\beta\omega(t)}-e^{-\gamma t-\beta\omega_{0}}\right].\label{eq:pe-LT}
\end{eqnarray}
Then the heat exchanged and the entropy change in the finite-time
isothermal process are 
\begin{eqnarray}
Q(t_{f}) & = & \omega(t_{f})p_{e}(t_{f})-\omega_{0}p_{e}(0)-\frac{1}{\beta}\ln\frac{1+e^{-\beta\omega_{0}}}{1+e^{-\beta\omega(t_{f})}}\nonumber \\
 &  & -\frac{\beta\epsilon^{2}e^{-\beta\omega_{0}}}{\gamma t_{f}-\beta\epsilon}\left(\frac{1-e^{-\beta\epsilon}}{\beta\epsilon}-\frac{1-e^{-\gamma t_{f}}}{\gamma t_{f}}\right)\label{eq:Q-LT}
\end{eqnarray}
and 
\begin{eqnarray}
\Delta S(t_{f}) & = & \beta\left[\omega(t_{f})p_{e}(t_{f})-\omega_{0}p_{e}(0)-\frac{1}{\beta}\ln\frac{1+e^{-\beta\omega_{0}}}{1+e^{-\beta\omega(t_{f})}}\right]\nonumber \\
 &  & -p_{e}(t_{f})\ln\left[1+\beta\epsilon\frac{e^{-\beta\omega(t_{f})}-e^{-\gamma t_{f}-\beta\omega_{0}}}{p_{e}^{0}(t_{f})(\gamma t_{f}-\beta\epsilon)}\right]\nonumber \\
 &  & -p_{g}(t_{f})\ln\left[1-\beta\epsilon\frac{e^{-\beta\omega(t_{f})}-e^{-\gamma t_{f}-\beta\omega_{0}}}{p_{g}^{0}(t_{f})(\gamma t_{f}-\beta\epsilon)}\right].\label{eq:S-LT}
\end{eqnarray}
The irreversible entropy production is straightforwardly obtained
as 
\begin{eqnarray}
\Delta S^{(i)}(t_{f}) & \approx & \frac{\left(\beta\epsilon\right)^{2}e^{-\beta\omega_{0}}}{\gamma t_{f}-\beta\epsilon}\left[\frac{1-e^{-\beta\epsilon}}{\beta\epsilon}-\frac{1-e^{-\gamma t_{f}}}{\gamma t_{f}}-\frac{\left(e^{-\beta\epsilon}-e^{-\gamma t_{f}}\right)^{2}}{p_{e}^{0}(t_{f})p_{g}^{0}(t_{f})}\frac{e^{-\beta\omega_{0}}}{\gamma t_{f}-\beta\epsilon}\right].\label{eq:Si-LT}
\end{eqnarray}
Similar as the high temperature case, the long-time behavior of $\Delta S^{(i)}$
is also of the $1/t_{f}$ form: 
\begin{equation}
\Delta S^{(i)}\left(\gamma t_{f}\gg\beta\epsilon\right)\approx\frac{\beta\epsilon}{\gamma t_{f}}e^{-\beta\omega_{0}}\left(1-e^{-\beta\epsilon}\right),\label{eq:LT-LT}
\end{equation}
and the short time limit is also finite: 
\begin{equation}
\Delta S^{(i)}(\gamma t_{f}\ll1)\approx e^{-\beta\omega_{0}}\left(\beta\epsilon+e^{-\beta\epsilon}-1\right).
\end{equation}
The low temperature irreversible entropy generation obtained by Eq.~(\ref{eq:LT-LT})
is well consistent with the numerical result, as shown in Fig.~\ref{fig:LowT}.

\begin{figure}
\includegraphics[width=8.5cm]{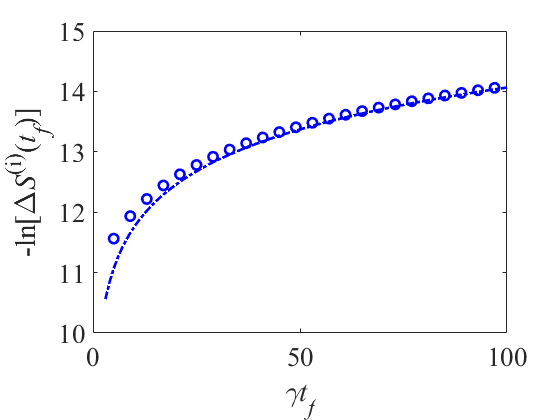}

\caption{\label{fig:LowT}Irreversible entropy generation as a function of
operation time at low temperature $\beta=1$. The circles are the
exact numerical result, and the dot-dashed line is the analytical
result obtained by Eq.~(\ref{eq:LT-LT}).}
\end{figure}
\end{widetext}

\end{document}